\documentclass[10pt,conference]{IEEEtran}

\usepackage{amsmath}
\usepackage{amssymb}
\usepackage{graphicx}
\usepackage{caption}
\usepackage{subcaption}
\usepackage{psfrag}
\usepackage{cite}

\newcommand{\bsx}{\boldsymbol{x}}

\newcommand{\bsy}{\boldsymbol{y}}

\newcommand{\bsw}{\boldsymbol{w}}
\newcommand{\bsz}{\boldsymbol{z}}

\newcommand{\bsI}{\boldsymbol{I}}
\newcommand{\bsO}{\boldsymbol{0}}
\newcommand{\Ebb}{\mathbb{E}}
\newcommand{\bstheta}{\boldsymbol{\theta}}
\newcommand{\bsphi}{\boldsymbol{\phi}}
\newcommand{\bspsi}{\boldsymbol{\psi}}
\newcommand{\bsPsi}{\boldsymbol{\Psi}}
\newcommand{\bsTheta}{\boldsymbol{\Theta}}
\newcommand{\bsPhi}{\boldsymbol{\Phi}}
\newcommand{\bsone}{\boldsymbol{1}}
\newcommand{\LR}{\textsf{LR}}
\newcommand{\LLR}{\textsf{LLR}}
\newcommand{\Define}{\stackrel{\Delta}{=}}
\newcommand{\argmax}{\arg\max}

\newcommand{\diag}{\text{diag}}
\newcommand{\diff}{\mathrm{d}}
\newcommand{\CN}{\mathcal{N}_{\mathbb{C}}}

\newcommand{\CircN}{\mathcal{VM}}
\newcommand{\VAR}{\mathsf{VAR}}
\newtheorem{myproposition}{\bf Proposition}
\newtheorem{mycorollary}{\bf Corollary}

\newtheorem{mylemma}{\bf Lemma}
\IEEEoverridecommandlockouts
\title{Optimal Detection in Training Assisted SIMO Systems with Phase Noise Impairments}
\author{\IEEEauthorblockN{Antonios Pitarokoilis, Emil Bj\"{o}rnson and Erik G. Larsson}
\IEEEauthorblockA{Department of Electrical Engineering, ISY, Link\"{o}ping University, 581 83 Link\"{o}ping, Sweden}
Email: \texttt{\{antonios.pitarokoilis,emil.bjornson,erik.g.larsson\}@liu.se} \thanks{This work was supported by the Swedish Foundation for Strategic Research (SSF) and ELLIIT.}}
\begin{document}

\maketitle

\begin{abstract}
In this paper, the problem of optimal maximum likelihood detection in a single user single-input multiple-output (SIMO) channel with phase noise at the receiver is considered. The optimal detection rules under training are derived for two operation modes, namely when the phase increments are fully correlated among the $M$ receiver antennas (synchronous operation) and when they are independent (non-synchronous operation). The phase noise increments are parameterized by a very general distribution, which includes the Wiener phase noise model as a special case. It is proven that phase noise creates a symbol-error-rate (SER) floor for both operation modes. In the synchronous operation this error floor is independent of $M$, while it goes to zero exponentially with $M$ in the non-synchronous operation.
\end{abstract}

\section{Introduction}\label{sec:Introduction}

The demand for wireless data traffic over cellular networks is expected to increase rapidly in the years to come. The deployment of base stations with an excess of base station (BS) antennas, $M$, \cite{Marzetta10}, termed as Massive multiple-input multiple-output (MIMO) systems, is a technology that offers unprecedented gains in energy and spectral efficiency \cite{Rusek13,CommMag13}. In particular, it has been shown that one can reduce the required radiated power by 1.5 dB to achieve a desired fixed information rate for every doubling of the base station (BS) antennas, when imperfect channel state information (CSI) is available and linear processing techniques are used \cite{Hien12TComm,JacobJSAC13}.

The gains offered by Massive MIMO can be achieved in the uplink by coherently combining the received signals using estimated channel impulse responses. A hardware impairment that hinders coherent communication is phase noise. Phase noise is introduced in communication systems by imperfections in the circuitry of the local oscillators that are used to convert the baseband signals to the passband and vice versa \cite{Demir00}. These imperfections are caused by the limited hardware complexity of practical circuit implementations. Since Massive MIMO relies on coherent combining, it is essential to study the effect of phase noise on the information rate of these systems.

Recent work has shown that in the phase noise impaired Massive MIMO uplink with imperfect CSI and linear receive processing, one can still reduce the radiated power by 1.5 dB for every doubling of $M$ and achieve a fixed desired per user information rate \cite{Allerton12,Asilomar13}. Further, it has been observed by various authors that using independent phase noise sources at the uplink receiver can provide improved performance. In \cite{Hoehne10} the authors show that the error vector magnitude (EVM) at the direction of the main lobe in beamforming is smaller with independent phase noise sources. In \cite{TWireless} a frequency selective Massive MIMO uplink with imperfect CSI and linear receive processing is considered. Lower bounds on the sum-rate performance for the cases of a common or many independent local oscillators are compared and it is shown that the sum-rate performance is better in the case of independent oscillators. Also, in \cite{DurisiSLOCLO} the authors find that the second order capacity expansion in the high signal-to-noise ratio (SNR) regime of a single-input multiple-output (SIMO) channel is higher for the case of separate oscillators. Furthermore, \cite{EmilTwireless14} shows that the phase noise variance can be allowed to increase logarithmically with $M$ in multi-cell systems with non-synchronous operation without losing much of the performance, while this is not possible in the synchronous operation. Finally, in \cite{EmilHardware14} the authors observe the same phenomenon for other types of hardware impairments.

In this work we consider the problem of optimal maximum likelihood (ML) detection in a single user SIMO channel with phase noise impairments at the receiver. Communication is done in two time slots: in the first time slot a training symbol is transmitted and in the second time slot a data symbol is sent. The receiver then calculates the ML detection rule using the information that has been acquired during training and data transmission. The ML detection rule is given explicitly for the cases of one common and multiple independent oscillators. The phase noise increments attain a very general parametrization, which enables the analysis of various phase noise models. Contrary to the no-phase-noise case, the high SNR analysis shows that the symbol-error-rate (SER) performance exhibits an error floor in both operations. For the synchronous operation, this error floor is independent of the number of receive antennas, $M$. However, for the non-synchronous operation, an upper bound on the SER shows that the error-floor is reduced to zero exponentially with $M$.

\section{System Model}\label{sec:SystemModel}

We consider a single antenna user that transmits information symbols to a BS array of $M$ antenna elements. The system is impaired by phase noise at the BS but not at the user terminal. Further, no fading is assumed. This assumption enables us to focus on the fundamental effects of phase noise. The combined effect of phase noise and channel fading will be considered in future work. Communication is scheduled in two time slots, each of duration $T_s$ seconds, where $T_s$ is the symbol interval. During the first time slot the user transmits a pilot symbol so that the receiver can get an initial phase estimate. The baseband equivalent representation of the received signal, $x_m$, at the $m$-th BS antenna at the first channel use is given by 
\begin{align}\label{eq:ReceivedTraining}
x_m=\sqrt{\rho}e^{j\theta_m} + w_m,
\end{align}
where $w_m$ is the $m$-th component of the additive white Gaussian noise (AWGN) column vector $\bsw\sim\CN\left(\bsO,\bsI_M\right)$ and is independent of $\theta_m$. The known positive scalar $\rho$ corresponds to the SNR measured at each received antenna. The phase $\theta_m$ is an unknown initial phase uniformly distributed in $[-\pi,\pi)$. In this work we consider two distinct operations. In the \emph{non-synchronous} operation, the phases $\theta_m$ are independent of each other. This corresponds to an operation where each antenna uses a separate oscillator for the downconversion of the bandpass signal to the baseband. In the \emph{synchronous} operation we assume that $\theta_1\equiv\cdots\equiv\theta_M\equiv\theta$, which corresponds to an operation where a common oscillator is used for the downconversion of the received passband signal to the baseband. 

During the second time slot the user transmits a data symbol, $s$, from a constellation, $\mathcal{S}$, with $\Ebb[s]=0$ and $\Ebb[|s|^2]=1$. The received signal at the $m$-th BS antenna is then given by
\begin{align}\label{eq:ReceivedData}
y_m=\sqrt{\rho}e^{j(\theta_m+\phi_m)}s+z_m,
\end{align}
%where $z_m$ is the $m$-th component of the additive white Gaussian noise column vector $\bsz\sim\CN\left(\bsO,\bsI_M\right)$, which is independent of $\theta_m$, $\phi_m$ and $s$. The additional phase drift, $\phi_m$ is assumed to be independent of $\theta_m$ and $s$. We assume that the oscillators used are free-running. A well-established model for free-running oscillators is the Wiener model \cite{Demir00}, where the discrete time phase noise process, $\vartheta[i]$, is given by
%\begin{align}\label{eq:WienerModel}
%\vartheta[i]=\vartheta[i-1] + \varphi.
%\end{align}
%The increment $\varphi$ is  independent of $\vartheta[i-1]$ and normally distributed $\mathcal{N}(0,\sigma_\varphi^2)$. 
where $z_m$ is the $m$-th component of the AWGN column vector $\bsz\sim\CN\left(\bsO,\bsI_M\right)$ and the vector $[\bsw^T~\bsz^T]$ is jointly Gaussian $\CN\left(\bsO,\bsI_{2M}\right)$. Also $\phi_m$ is the random phase noise increment, independent of $\theta_m$, the information symbol, $s$, and the AWGN noise, $z_m$. Since we have $\phi_m\in[-\pi,\pi)$, the probability density function (pdf), $p_{\Phi_M}(\phi_m)$, of the phase noise increment $\phi_m$ can be expressed by its Fourier expansion \cite{mardia2009directional,CircularStatisticsBook}. In this work we consider a general model for $p_{\Phi_M}(\phi_m)$, i.e.
\begin{align}\label{eq:PhaseNoiseIncrement}
p_{\Phi_m}(\phi_m)=\frac{1}{2\pi}\left(\alpha_{m,0}+2\sum_{l=1}^{\infty}\alpha_{m,l}\cos\left(l\phi_m\right)\right),
\end{align}
where $\alpha_{m,l},~l=0,1,2,\ldots$, are real and known constants. The Fourier expansion in \eqref{eq:PhaseNoiseIncrement} can model any continuous, differentiable, unimodal, even and zero mean pdf. Later in the paper we will select a specific pdf, namely, the circular normal distribution, however, most of the results presented here are valid for any choice of $p_{\Phi_m}(\phi_m)$ that has an expansion as in \eqref{eq:PhaseNoiseIncrement}.\footnote{The circular normal distribution is often used in literature to model the phase disturbance at the output of a local oscillator \cite{Demir00}.} For the synchronous operation the phase noise increments $\phi_m$ are assumed to be identical among the antennas, whereas for the non-synchronous operation they are assumed independent. Finally, the system model for the non-synchronous operation is given in matrix vector form by
\begin{align}
\bsx &= \sqrt{\rho}\bsTheta\bsone + \bsw\label{eq:MatVecReceivedPilot}\\
\bsy &= \sqrt{\rho}\bsTheta\bsPhi s\bsone + \bsz\label{eq:MatVecReceivedData},
\end{align}
where $\bsone$ is the all-one column vector of size $M$, $\bsTheta\Define\diag\left\{e^{j\theta_1},\ldots,e^{j\theta_M}\right\}$ and $\bsPhi\Define\diag\left\{e^{j\phi_1},\ldots,e^{j\phi_M}\right\}$. The system model for the synchronous operation is trivially derived from \eqref{eq:MatVecReceivedPilot} and \eqref{eq:MatVecReceivedData} by setting $\bsTheta=e^{j\theta}\bsI_M$ and $\bsPhi=e^{j\phi}\bsI_M$.

\section{Maximum Likelihood Detection}\label{sec:MLDetection}

In this section we derive the likelihood function of the received vectors $\bsx$ and $\bsy$ given the transmitted symbol, $s$, $p(\bsx,\bsy|s)$, for both operations. Then the ML detection rule is given by
\begin{align}\label{eq:MLruleDefinition}
\hat s=\argmax_{s\in\mathcal{S}}p(\bsx,\bsy|s).
\end{align}
\begin{myproposition}\label{prop:LikelihoodFunctionNonSynch}
The pdf of the received vectors $(\bsx,\bsy)$ given a symbol $s$ for the non-synchronous operation is given by
\begin{align}\label{eq:LikelihoodNonSynch}
p(\bsx,\bsy|s)&=A\prod_{m=1}^{M}\left(\beta_{m,0}+2\sum_{l=1}^{\infty}\beta_{m,l}\cos\left(l\zeta_m\right)\right)
\end{align}
where 
\begin{align}
A&\Define\exp\left(-\|\bsx\|^2-\|\bsy\|^2-\rho M(1+|s|^2)\right)/\pi^{2M},\\
\beta_{m,l}&\Define\alpha_{m,l} I_l(2\sqrt{\rho}|s^*y_m|)I_l(2\sqrt{\rho}|x_m|),\\
\zeta_m&\Define\arg(y_m)-\arg(x_m)-\arg(s).
\end{align}
$I_l(\cdot)$ is the $l$-th order modified Bessel function of first kind\cite{VanTreesPartI} and $\|\cdot\|$ is the Euclidean norm.
\end{myproposition}
\begin{IEEEproof}
Define the vectors $\bstheta\Define[\theta_1,\ldots,\theta_M]^T$ and $\bsphi\Define[\phi_1,\ldots,\phi_M]^T$. The likelihood function is given by
\begin{align}\label{eq:LikelihoodNonSynchFactorization}
&p(\bsx,\bsy|s)=\iint p(\bsx,\bsy|s,\bstheta,\bsphi)p(\bstheta,\bsphi|s)\diff  \bstheta \diff \bsphi\nonumber\\
&\stackrel{(a)}{=}\iint p(\bsx|\bstheta)p(\bsy|s,\bstheta,\bsphi)p(\bstheta)p(\bsphi)\diff \bstheta \diff \bsphi\nonumber\\
%\end{align*}
%\begin{align*}
&\stackrel{(b)}{=}\!\!\prod_{m=1}^{M}\!\int_{-\pi}^{\pi}\!\!\!\!\!p(x_m|\theta_m)p(\theta_m)\underbrace{\int_{-\pi}^{\pi}\!\!\!\!\!p(y_m|s,\theta_m,\phi_m)p(\phi_m)\diff \phi_m}_{\Define \mathcal{I}_m(y_m|\theta_m,s)} \diff \theta_m,
%&=\prod_{m=1}^{M}\int_{-\pi}^{\pi}p(x_m|\theta_m)p(\theta_m)A_m(\theta_m|s) \diff \theta_m
\end{align}
where (a) follows from the fact that conditioned on $\bstheta,~\bsphi$ and $s$, the vectors $\bsx$ and $\bsy$ are independent and (b) is a consequence of the independence of the components in $\bstheta$, $\bsphi$, $\bsw$ and $\bsz$. The channel probability law $p(y_m|s,\theta_m,\phi_m)$ can be expressed as 
\begin{align}\label{eq:InnerChannelLaw}
&p(y_m|s,\theta_m,\phi_m)=\frac{1}{\pi}\exp\left(-\left|y_m-\sqrt{\rho}e^{j(\theta_m+\phi_m)}s\right|^2\right)\nonumber\\
%&=\frac{e^{-|y_m|^2-\rho |s|^2}}{\pi}\exp\left(\Re\left\{2\sqrt{\rho}s^*y_m e^{-j\theta_m}e^{-j\phi_m}\right\}\right)\\
&=\frac{e^{-|y_m|^2-\rho |s|^2}}{\pi}\exp\left(2\sqrt{\rho}|s^*y_m|\cos\left(\phi_m\!+\!\theta_m\!-\!\arg\left\{s^*y_m\right\}\right)\right)\nonumber\\
&=\frac{e^{-|y_m|^2-\rho |s|^2}}{\pi}\Bigg(I_0(2\sqrt{\rho}|s^*y_m|)\\
&\left.+2\sum_{l=1}^{\infty}I_l(2\sqrt{\rho}|s^*y_m|)\cos\left(l\left(\phi_m+\theta_m-\arg\left(s^*y_m\right)\right)\right)\right),\nonumber
\end{align}
where the last step follows from the Jacobi-Anger formula
\begin{align}\label{eq:JacobiAnger}
e^{\alpha\cos\beta}=I_0(\alpha)+2\sum_{l=1}^{\infty}I_l(\alpha)\cos(l\beta).
\end{align}
Then, the integral $\mathcal{I}_m(y_m|\theta_m,s)$ is
\begin{align*}
\mathcal{I}_m&(y_m|\theta_m,s)=\int_{-\pi}^{\pi}p(y_m|s,\theta_m,\phi_m)p(\phi_m)\diff \phi_m\\
&=\frac{e^{-|y_m|^2-\rho |s|^2}}{\pi}\Bigg(\alpha_{m,0} I_0(2\sqrt{\rho}|s^*y_m|)\\
&\left.+2\sum_{l=1}^{\infty}\alpha_{m,l} I_l(2\sqrt{\rho}|s^*y_m|)\cos\left(l\left(\theta_m-\arg\left(s^*y_m\right)\right)\right)\right).
\end{align*}
By manipulating $p(x_m|\theta_m)$ in the same way as in \eqref{eq:InnerChannelLaw} and by the orthogonality of the trigonometric functions we obtain
\begin{align}\label{eq:partialLikelihoodsNonSynch}
&p(x_m,y_m|s)=\int_{-\pi}^{\pi}p(x_m|\theta_m)p(\theta_m)\mathcal{I}_m(y_m|\theta_m,s) \diff \theta_m\nonumber\\
&=\frac{\exp\left(-|x_m|^2-|y_m|^2-\rho (1+|s|^2)\right)}{\pi^2}\nonumber\\
&\times\left(\beta_{m,0}+2\sum_{l=1}^{\infty}\beta_{m,l}\cos\left(l\zeta_m\right)\right).
\end{align}
The result in \eqref{eq:LikelihoodNonSynch} follows by substituting \eqref{eq:partialLikelihoodsNonSynch} in \eqref{eq:LikelihoodNonSynchFactorization}.
\end{IEEEproof}
The expression in \eqref{eq:LikelihoodNonSynch} holds for any constellation, but can be particularized for any choice of constellation. In the following, we particularize \eqref{eq:LikelihoodNonSynch} for the phase shift keying constellation with $N$ symbols ($N$-PSK). This choice will be motivated in Section \ref{sub:HighSNRNonSynch}.
\begin{mycorollary}\label{cor:PSK_LLR_nsynch}
For $s$ selected from an $N$-PSK constellation as $s\in\left\{1,e^{j\frac{2\pi}{N}},\ldots,e^{j\frac{2\pi(N-1)}{N}}\right\}$, the likelihood \eqref{eq:LikelihoodNonSynch} can be written as
\begin{align}\label{eq:LikelihoodNonSynchPSK}
p\left(\bsx,\bsy|s=e^{j\frac{2\pi n}{N}}\right)=A\prod_{m=1}^{M}f_{m,n}(\arg(y_m)-\arg(x_m))
\end{align}
where $f_{m,n}(\omega_m)\Define\sum_{l=-\infty}^{\infty}\beta_{m,|l|}e^{jl\left(\omega_m-\frac{2\pi n}{N}\right)}.$
%\begin{align*}
%
%\end{align*}
The log-likelihood function for the symbol $e^{j\frac{2\pi n}{N}}$ is given by
\begin{align}\label{eq:PSKnonSynchLLR}
\LR_n\Define\sum_{m=1}^{M}\ln\left(f_{m,0}(\arg(y_m)-\arg(x_m)-\frac{2\pi n}{N})\right).
\end{align}
\end{mycorollary}
\begin{IEEEproof}
Eq. \eqref{eq:LikelihoodNonSynchPSK} follows immediately for $|s|=1$. Eq. \eqref{eq:PSKnonSynchLLR} follows by observing that $f_{m,n}(\omega_m)=f_{m,0}\left(\omega_m-\frac{2\pi n}{N}\right)$.
\end{IEEEproof}
Next, we derive the counterparts of Proposition \ref{prop:LikelihoodFunctionNonSynch} and Corollary \ref{cor:PSK_LLR_nsynch} for the synchronous operation.
\begin{myproposition}\label{prop:LikelihoodFunctionSynch}
The pdf of the received vectors $(\bsx,\bsy)$ given a symbol $s$ for the synchronous operation is given by
\begin{align}\label{eq:LikelihoodSynch}
p(\bsx,\bsy|s)&=A\left(\beta_0+2\sum_{l=1}^{\infty}\beta_l\cos\left(l\zeta\right)\right)
\end{align}
where 
\begin{align}
\beta_l&\Define\alpha_l I_l(2\sqrt{\rho}|s^*\bsone^T\bsy|)I_l(2\sqrt{\rho}|\bsone^T\bsx|),\\
\zeta&\Define\arg(\bsone^T\bsy)-\arg(\bsone^T\bsx)-\arg(s).
\end{align}
\end{myproposition}
\begin{IEEEproof}
The likelihood function in this case is given by
\begin{align*}
&p(\bsx,\bsy|s)=\int p(\bsx|\theta)p(\theta)\int p(\bsy|s,\theta,\phi)p(\phi)\diff \phi\diff \theta.
\end{align*}
The proof follows steps that are similar to Proposition \ref{prop:LikelihoodFunctionNonSynch} with the observation that the channel laws $p(\bsx|\theta)$ and $p(\bsy|s,\theta,\phi)$ are $M$-variate complex Gaussian distributions.
\end{IEEEproof}
\begin{mycorollary}\label{cor:PSK_LR_synch}
For $s$ selected from an $N$-PSK constellation and $|s|=1$, the log-likelihood function for the symbol $e^{j\frac{2\pi n}{N}}$ can be written as
\begin{align}\label{eq:PSKSynchLR}
\LR_n\!=\!\sum_{l=1}^{\infty}\!\beta_l\sin\left(\!\frac{l\pi n}{N}\!\right)\!\sin\left(\!l\!\left(\arg(\bsone^T\bsy)\!-\!\arg(\bsone^T\bsx)\!-\!\frac{\pi n}{N}\right)\right).
\end{align}
\end{mycorollary}

\subsection{High SNR Analysis for the Synchronous Operation}\label{sub:HighSNRSynch}

The expressions in Propositions \ref{prop:LikelihoodFunctionNonSynch}, \ref{prop:LikelihoodFunctionSynch} and Corollaries \ref{cor:PSK_LLR_nsynch} and \ref{cor:PSK_LR_synch} can be easily implemented\footnote{For a fixed argument, $x$, the value of $I_\mu(x)$ reduces rapidly as $\mu$ increases. Hence, only a small number of terms is required to approximate accurately the detectors in Propositions \ref{prop:LikelihoodFunctionNonSynch} and \ref{prop:LikelihoodFunctionSynch} \cite{Colavolpe05JSAC}.} but do not allow for analytical work. Therefore, in this section we present an asymptotic analysis as $\rho\rightarrow\infty$ (high SNR) for the problem in \eqref{eq:MatVecReceivedPilot} and \eqref{eq:MatVecReceivedData}. We start with the synchronous operation as it appears to be simpler.
The system model for the synchronous operation in the case of high SNR can be expressed as
\begin{align}\label{eq:highSNRsynch}
\tilde\bsx&\Define\frac{\bsx}{\sqrt{\rho}}=e^{j\theta}\bsone\nonumber\\
\!\tilde\bsy\!\Define\!\frac{\bsy}{\sqrt{\rho}}\!=\!e^{j(\theta+\phi)}s\bsone\!\Rightarrow\!\tilde\bsy\!&=\!\tilde\bsx e^{j\phi}s\!\Rightarrow\!\begin{cases}
|\tilde\bsx^H\tilde\bsy|\!=\!M|s|\\
\psi\!=\!\arg(s)\!+\!\phi 
\end{cases}
\end{align}
where $\psi\Define\arg\left(\tilde\bsx^H\tilde\bsy\right)$. From \eqref{eq:highSNRsynch} it is apparent that the amplitude of $s$ can be decoded error-free. Hence we restrict the study to PSK constellations where the signal is impaired by phase noise. The observation $\psi$ is in this case sufficient statistics. We proceed by deriving the asymptotic SER at high SNR for PSK constellations. We further specify the pdf in \eqref{eq:PhaseNoiseIncrement} to be the circular normal (also known as von Mises) distribution with zero mean and concentration parameter $\kappa\geq 0$, $\phi\sim\CircN(0,\kappa)$, which is given by
\begin{align}\label{eq:vonMises}
p_\Phi(\phi)&=\frac{e^{\kappa\cos\phi}}{2\pi I_0(\kappa)}=\frac{1}{2\pi}\left(1+2\sum_{l=1}^{\infty}\frac{I_l(\kappa)}{I_0(\kappa)}\cos\left(l\phi\right)\right),
\end{align}
for $\phi\in[-\pi,\pi)$. Then $\psi|\arg(s)\sim\CircN(\arg(s),\kappa)$. By the symmetry of the circular normal distribution around its mean, the symmetry of the PSK constellations and the fact that we assume equal priors for the constellation symbols, the decision region for the symbol $s_n=\exp\left(j\frac{2\pi n}{N}\right),~n=0,\ldots,N-1$ is $\left[\frac{2\pi n}{N}-\frac{\pi}{N},\frac{2\pi n}{N}+\frac{\pi}{N}\right)$. Then, the probability of error at high SNR is given by
\begin{align}\label{eq:ErrorProbabilityHighSNRSynch}
\Pr\left\{\epsilon\right\}&\Define 1-\int_{-\frac{\pi}{N}}^{\frac{\pi}{N}}p_\Phi(\phi)\diff\phi=1-\int_{-\frac{\pi}{N}}^{\frac{\pi}{N}}\frac{e^{\kappa\cos\phi}}{2\pi I_0(\kappa)}\diff\phi.
\end{align}
\begin{myproposition}\label{prop:highSNRSynch}
From \eqref{eq:ErrorProbabilityHighSNRSynch} we observe that there is a non-zero SER floor for the synchronous operation, which depends only on the concentration parameter of the phase noise increment, $\kappa$, and the PSK constellation density, $N$, but is independent of the number of receive antennas, $M$.
\end{myproposition}

\subsection{High SNR Analysis for the Non-Synchronous Operation}\label{sub:HighSNRNonSynch}

The system model \eqref{eq:ReceivedTraining} and \eqref{eq:ReceivedData} for the non-synchronous operation in the high SNR regime is given by
\begin{align}\label{eq:SystemModelHighSNRNonSynch}
\tilde x_m\Define\frac{x_m}{\sqrt{\rho}}&=e^{j\theta_m}\nonumber\\
\tilde y_m\Define\frac{y_m}{\sqrt{\rho}}&=e^{j(\theta_m+\phi_m)}s\Rightarrow\nonumber\\
\psi_m&\Define\arg(\tilde x_m^*\tilde y_m) = \phi_m + \arg(s).
\end{align}
Similarly to \eqref{eq:highSNRsynch}, the amplitude can be decoded error free also in the non-synchronous operation. Hence, we focus only on the phase. The likelihood function under the assumption that $\phi_m$ are independent $\CircN(0,\kappa)$ random variables is given by
\begin{align*}
p_{\bsPsi|s}(\bspsi|s)=\prod_{m=1}^{M}p_{\Psi_m|s}(\psi_m|s)=\frac{e^{\kappa\sum_{m=1}^{M}\cos\left(\psi_m-\arg(s)\right)}}{\left(2\pi I_0(\kappa)\right)^M}.
\end{align*}
\begin{myproposition}\label{prop:highSNRLLRNonSynch}
The scaled log-likelihood ratio for the symbol $e^{j\frac{2\pi n}{N}}$ from an $N$-PSK constellation is given by
\begin{align}\label{eq:highSNRLLRNonSynch}
\LLR_n&\Define\frac{1}{2\kappa M}\ln\left(\frac{p_{\bsPsi|s}(\bsPsi|\arg(s)=\frac{2\pi n}{N})}{p_{\bsPsi|s}(\bsPsi|\arg(s)=0)}\right)\nonumber\\
&=\frac{1}{M}\sum_{m=1}^{M}\sin\left(\frac{\pi n}{N}\right)\sin\left(\psi_m-\frac{\pi n}{N}\right).
\end{align}
If we denote by $\epsilon$ the error event, i.e. the case where the detected symbol $\hat s$ is different from the transmitted symbol $s$, then the symbol error probability is given by
\begin{align}\label{eq:highSNRSERNonSynch}
\Pr\left\{\epsilon\right\}&= \Pr\left\{\bigcup_{n=1}^{N-1}\left\{\LLR_n>0\right\}\middle |\arg(s)=0\right\}.%\Pr\left\{\epsilon\middle |\arg(s)=0\right\}=1-\Pr\left\{\bigcap_{n=1}^{N-1}\left\{\LLR_n<0\right\}\middle |\arg(s)=0\right\}
\end{align}
\end{myproposition}
The exact calculation of the probability of error in \eqref{eq:highSNRSERNonSynch} appears formidable. We therefore derive an upper bound on the pairwise symbol error probability of erroneously detecting $s_n=\exp\left(j\frac{2\pi n}{N}\right),~n=1,\ldots,N-1$ when $s_0=1$ was sent\footnote{We note that due to the symmetry of the von Mises distribution around its mean and the uniform priors on the input symbols, the conditioning on any particular input symbol does not affect the result. The symbol $s_0=1$ with $\arg(s_0)=0$ is selected for convenience.}. For this purpose we will use the result stated in the following lemma.
\begin{mylemma}{\emph{Bernstein Inequality}\cite{Bernstein}}\label{lem:Bernstein}
Let $X_m,~m=1,\ldots,M$ be independent and identically distributed random variables with $\Ebb [X_m]=0$, $|X_m|<C$ almost surely for some bounded constant $C$, $X_s\Define\sum_{m=1}^{M}X_m$ and $\varsigma\Define\sqrt{\VAR(X_s)}$. Then for all $t>0$
\begin{align*}
\Pr\left\{X_s>t\varsigma\right\}\leq\exp\left(-\frac{t^2}{2+\frac{2}{3}\frac{C}{\varsigma}t}\right).
\end{align*}
\end{mylemma}
\begin{myproposition}\label{prop:Bernstein}
The pairwise error probability for the detected symbol $\hat s_n$ to be $s_n=\exp\left(j\frac{2\pi n}{N}\right),~n=1,\ldots,N-1$ given that the symbol $s_0=1$ was sent is upper bounded by
\begin{align}\label{eq:BernsteinBound}
\Pr\left\{\LLR_n>0\right\}\leq\exp\left(-\frac{M\left(\frac{\sin^2\left(\frac{\pi n}{N}\right)}{\sqrt{\VAR(X_{m,n})}}\frac{I_1(\kappa)}{I_0(\kappa)}\right)^2}{2+\frac{2}{3}\frac{C\sin^2\left(\frac{\pi n}{N}\right)}{\VAR(X_{m,n})}\frac{I_1(\kappa)}{I_0(\kappa)}}\right),
\end{align}
where $$C\Define\sin\left(\frac{\pi n}{N}\right)+\sin^2\left(\frac{\pi n}{N}\right)\frac{I_1(\kappa)}{I_0(\kappa)}$$ and 
\small
\begin{align}\label{eq:VarianceXmn}
&\VAR(X_{m,n})\\
&=\sin^2\left(\frac{\pi n}{N}\right)\left(\frac{I_1(\kappa)\cos\left(\frac{2\pi n}{N}\right)}{\kappa I_0(\kappa)}+\sin^2\left(\frac{\pi n}{N}\right)\left(1-\frac{I_1^2(\kappa)}{I_0^2(\kappa)}\right)\right).\nonumber
\end{align}\normalsize
\end{myproposition}
\begin{IEEEproof}
Let \[X_{m,n}\Define\sin\left(\frac{\pi n}{N}\right)\sin\left(\psi_m-\frac{\pi n}{N}\right)+\sin^2\left(\frac{\pi n}{N}\right)\frac{I_1(\kappa)}{I_0(\kappa)}.\]
Then $\Ebb [X_{m,n}]=0$, $|X_{m,n}|\leq C\Define\sin\left(\frac{\pi n}{N}\right)+\sin^2\left(\frac{\pi n}{N}\right)\frac{I_1(\kappa)}{I_0(\kappa)}$ and $\VAR(X_{m,n})$ is given by \eqref{eq:VarianceXmn}. Then
\begin{align*}
&\Pr\left\{\LLR_n>0\right\}\\
&=\Pr\left\{\frac{1}{M}\sum_{m=1}^{M}\sin\left(\frac{\pi n}{N}\right)\sin\left(\psi_m-\frac{\pi n}{N}\right)>0\right\}\\
&=\Pr\left\{\sum_{m=1}^{M}X_{m,n}>M\sin^2\left(\frac{\pi n}{N}\right)\frac{I_1(\kappa)}{I_0(\kappa)}\right\}.
\end{align*}
Define $\varsigma\Define\sqrt{M}\sqrt{\VAR(X_{m,n})}$ and \[t\Define\sqrt{M}\frac{\sin^2\left(\frac{\pi n}{N}\right)}{\sqrt{\VAR(X_{m,n})}}\frac{I_1(\kappa)}{I_0(\kappa)}.\]
By applying the Bernstein inequality (Lemma \ref{lem:Bernstein})
\begin{align*}
\Pr\left\{\LLR_n>0\right\}\leq \exp\left(-\frac{M\left(\frac{\sin^2\left(\frac{\pi n}{N}\right)}{\sqrt{\VAR(X_{m,n})}}\frac{I_1(\kappa)}{I_0(\kappa)}\right)^2}{2+\frac{2}{3}\frac{C\sin^2\left(\frac{\pi n}{N}\right)}{\VAR(X_{m,n})}\frac{I_1(\kappa)}{I_0(\kappa)}}\right).
\end{align*}
\end{IEEEproof}
\begin{mycorollary}\label{cor:LargeMProbabilityOfError}
The pairwise error probability goes to zero at least exponentially with $M$ for the non-synchronous operation. Also for the SER in \eqref{eq:highSNRSERNonSynch} we have $\lim_{M\rightarrow\infty}\Pr\left\{\epsilon\right\}\rightarrow 0$.
\end{mycorollary}
\begin{IEEEproof}
We observe that $C$ and $\VAR(X_{m,n})$ are constant with respect to $M$. Hence, the bound in \eqref{eq:BernsteinBound} behaves as $O(e^{-M})$. Further, using the union bound\cite{proakis} and Proposition \ref{prop:Bernstein}, the SER can be upper bounded by
\begin{align*}
\lim_{M\rightarrow\infty}\Pr\left\{\epsilon\right\}&\leq\sum_{n=1}^{N} \lim_{M\rightarrow\infty}\Pr\left\{\LLR_n>0\right\}=0,
\end{align*}
which establishes the second claim of Corollary \ref{cor:LargeMProbabilityOfError}.
\end{IEEEproof}

Hence, in the non-synchronous case the error floor can be made arbitrarily small by increasing the number of receive antennas. This is in contrast to the synchronous case, shown in Proposition \ref{prop:highSNRSynch}, where an irreducible SER floor independent of $M$ was observed.

\section{Numerical Examples}\label{sec:NumericalExamples}

\begin{figure}
\includegraphics[width=0.47\textwidth]{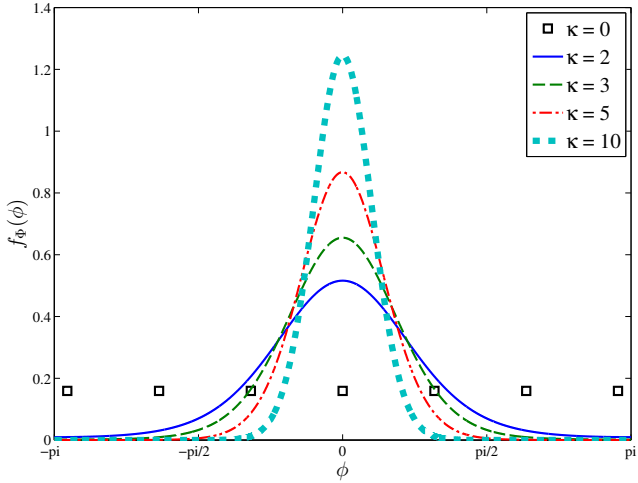}
\caption{The zero mean circular normal (von Mises) distribution \eqref{eq:vonMises} for various values of the concentration parameter $\kappa\in\{0, \, 2, \, 3, \, 5, \, 10\}$. For $\kappa=0$ the distribution is uniform in $[-\pi,\pi)$ and becomes more concentrated around the mean as $\kappa$ increases.}
\label{fig:vonMises}
\end{figure}

\begin{figure}
\includegraphics[width=0.47\textwidth]{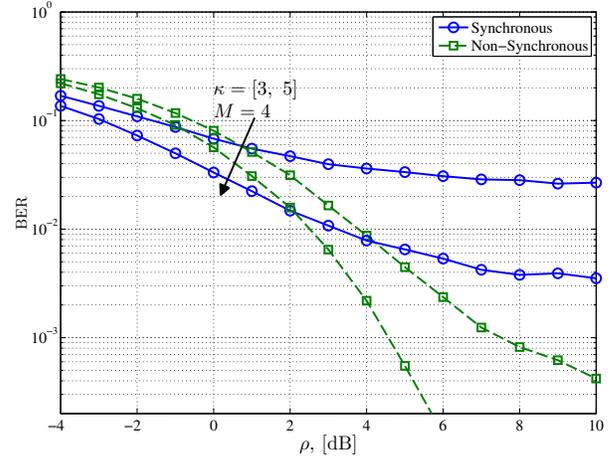}
\caption{Bit-Error-Rate (BER) performance of BPSK modulation as a function of $\rho$ [dB] for fixed $M = 4$ and two choices of $\kappa$, i.e. $\kappa = 3$ and $\kappa = 5$.}
\label{fig:TrainingBasedBPSK}
\end{figure}

In this section we present numerical examples based on the theoretical results presented in Section \ref{sec:MLDetection}. The phase noise increments, $\phi_m,~m=1,\ldots,M$, for the non-synchronous case and $\phi$ for the synchronous case are distributed as $\CircN(0,\kappa)$, where $\kappa$ is specified for each figure. In Fig. \ref{fig:vonMises} the zero-mean circular normal distribution is plotted for various values of $\kappa\in\{0, \, 2, \, 3, \, 5, \, 10\}$. For $\kappa=0$ the distribution is uniform in $[-\pi,\pi)$, whereas it becomes more concentrated around the mean as $\kappa$ increases. In Fig. \ref{fig:TrainingBasedBPSK} the bit-error-rate (BER) performance of the synchronous and non-synchronous operation for a BPSK constellation\footnote{For BPSK modulation one symbol is one bit, so the BER is equal to the SER.} is calculated via Monte-Carlo simulations based on the detectors in Propositions \ref{prop:LikelihoodFunctionNonSynch} and \ref{prop:LikelihoodFunctionSynch} and plotted as a function of $\rho$ for a fixed number of BS antennas, $M=4$. %The decision rules for the detection of the bits are given by \eqref{eq:PSKSynchLR} and \eqref{eq:PSKnonSynchLLR} for the synchronous and non-synchronous case respectively.
Two families of curves are plotted, namely for $\kappa=3$ and for $\kappa=5$. We observe that in the high SNR regime the non-synchronous operation exhibits better BER performance. The BER floor that appears for the synchronous operation depends on the concentration parameter of the phase noise increment, $\kappa$, and is reduced as $\kappa$ increases, which is in accordance with Proposition \ref{prop:highSNRSynch}.

%\begin{figure}
%\includegraphics[width=0.47\textwidth]{HighSNRErrProbSynchSemiLog}
%\caption{High SNR error probability for Synchronous Operation, with various choices of the phase noise increment concentration parameter.}
%\label{fig:HighSNRErrProbSynchSemiLog}
%\end{figure}

\begin{figure}
\includegraphics[width=0.47\textwidth]{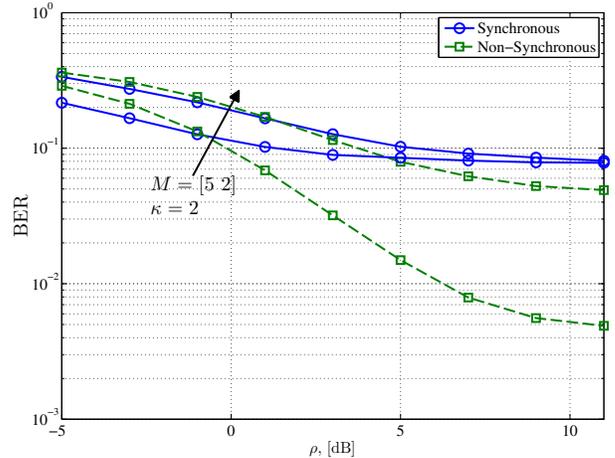}
\caption{Bit-Error-Rate (BER) performance of BPSK modulation as a function of $\rho$ [dB] for fixed $\kappa = 2$ and two choices of $M$, i.e. $M = 2$ and $M = 5$.}
\label{fig:BPSKberK2varM}
\end{figure}

\begin{figure}
\includegraphics[width=0.47\textwidth]{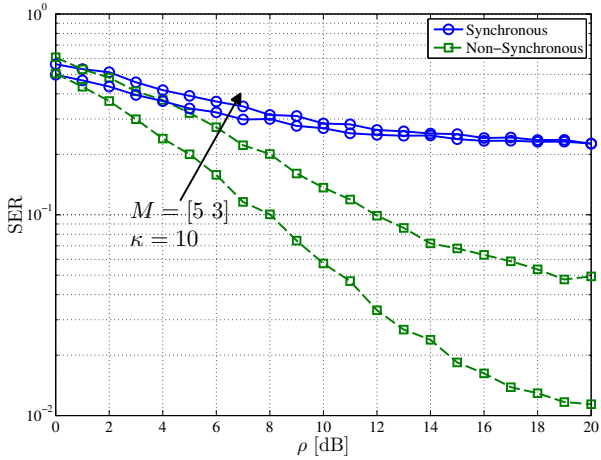}
\caption{Symbol-Error-Rate (SER) performance of 8-PSK modulation as a function of $\rho$ [dB] for fixed $\kappa = 10$ and two choices of $M$, i.e. $M = 3$ and $M = 5$.}
\label{fig:ser8PSKvarM}
\end{figure}

In Fig. \ref{fig:BPSKberK2varM} the BER performance of the synchronous and non-synchronous operation with BPSK symbols is plotted as a function of $\rho$ for fixed $\kappa=2$. Two families of curves are shown, namely for $M=2$ and for $M=5$. The BER performance in the high SNR is superior in the non-synchronous case. As in Fig. \ref{fig:TrainingBasedBPSK}, an error floor is observed for the synchronous operation. However, the error floor appears to be independent of $M$, as noted in Proposition \ref{prop:highSNRSynch}. The error floor appears also in the non-synchronous operation. However, it is lower than the respective floor of the synchronous case. In addition, it is reduced as $M$ increases, as proved in Proposition \ref{prop:highSNRLLRNonSynch}. Similar behavior has been observed in \cite{Hoehne10,TWireless,DurisiSLOCLO,EmilTwireless14,EmilHardware14}. Finally, in Fig. \ref{fig:ser8PSKvarM} the SER of an 8-PSK system is plotted as a function of $\rho$ for fixed $\kappa=10$ and two values of $M$, $M=3$ and $M=5$. Fig. \ref{fig:ser8PSKvarM} extends the conclusions of Fig. \ref{fig:BPSKberK2varM} to a multi-symbol constellation and verifies Propositions \ref{prop:highSNRSynch} and \ref{prop:highSNRLLRNonSynch}.

We conclude the discussion on the numerical examples by noting that the synchronous operation shows better performance at low SNR under certain circumstances. This is particularly the case in Fig. \ref{fig:TrainingBasedBPSK} for the curve corresponding to $M=4$ and $\kappa=5$. On the other hand, for 8-PSK constellation both operations exhibit similar performance, as shown in Fig. \ref{fig:ser8PSKvarM}. In the non-synchronous operation the independent phase noise components have an effect that is similar to the additive noise so this component is amplified. However, in the synchronous operation phase noise is a common rotation of the signal at all antenna elements. When the variance of the phase noise increments is sufficiently small and the constellation is not dense, then the common phase rotation due to phase noise degrades the SER performance more moderately in comparison to the independent rotations that appear in the non-synchronous operation. However, when the constellation is dense, i.e. sensitive to common rotations, this advantage of the synchronous operation disappears. This is the reason why we observe similar SER performance in the low SNR regime of Fig. \ref{fig:ser8PSKvarM} for the case of 8-PSK. It is clear that there is a regime where the synchronous operation is better but after a transition region the non-synchronous operation exhibits superior performance. The precise calculation of this transition region appears to be dependent on many parameters such as the SNR, the constellation density and the variance of the phase noise innovations, however, it will be considered in future work.

\section{Conclusions}\label{sec:Conclusions}

In this paper, the problem of optimal ML detection in a phase noise impaired SIMO systems with training was considered. Two different operation modes were investigated, namely the case of a common oscillator for the whole receive array (synchronous operation) and the case of separate oscillators for each receive antenna element (non-synchronous operation). The optimal ML detection rules were derived under a very general parameterization for the phase noise increment distribution. For both operations, the asymptotic ML detection rules were derived in the high SNR regime when the phase noise increments were distributed according to a circular normal (von Mises) distribution. An SER floor was observed for both operations. The exact expression of this floor was derived for the synchronous operation, which showed that it depends only on the constellation density and the concentration parameter of the von Mises distribution but is independent of the number of BS antenna elements. An upper bound on the pairwise error probability was derived for the non-synchronous operation. This bound was shown to approach zero exponentially in $M$, which implies that the probability of error can be made arbitrarily small by increasing the number of BS antennas. Numerical examples were shown to support the derived propositions.

\bibliographystyle{ieeetr}
\bibliography{phNbib}

\end{document}